\documentclass[seceq]{ptptex}
\usepackage{wrapft}

\usepackage{amssymb}
\usepackage[dvips]{graphicx}
\usepackage{booktabs}
\usepackage{tabularx}

\title{Nonequilibrium Steady States and MacLennan-Zubarev Ensembles
in a Quantum Junction System
}

\author{
Shuichi \textsc{Tasaki}, Junko \textsc{Takahashi}
}

\inst{
Advanced Institute for Complex Systems
and 
Department of Applied Physics,\\
School
of Science and Engineerings,
Waseda University,\\
3-4-1 Okubo, Shinjuku-ku,
Tokyo 169-8555,
Japan
}

\abst{
Based on a recent progress in nonequilibrium statistical mechanics
of infinitely extended quantum systems, a nonequlibrium steady state
(NESS) is constructed for a single-level quantum dot interacting with 
two free reservoirs under less general but more practically useful 
conditions than the previous works. 
As an example, a model of an Ahoronov-Bohm ring with a quantum dot
is studied in detail.
Then, NESS is shown to be regarded as a MacLennan-Zubarev 
ensemble. A formal relation between response 
and correlation at NESS is derived as well.
}

\begin{document}
\maketitle

\section{Introduction}\label{Intro}
Nonequilibrium statistical ensembles have been studied for many years, but
no consensus has been made. As an illustration, let us consider a 
MacLennan-Zubarev 
ensemble for a system consisting of identical particles. The system is 
assumed to 
be divided into $M$ independent parts, each of which has the energy 
$H_j$ and the 
particle number $N_j$ ($j=1,\cdots M$) and which are interacting by 
an interaction $W$. 
According to the MacLennan-Zubarev approach\cite{MacLennan,Zubarev}, 
a steady state close to a local equilirium state is described by:
\begin{equation}
\rho_+ = {e^{-\sum_{j=1}^M \beta_j \bigl({\widetilde H}_j - \mu_j
{\widetilde N}_j \bigr)}\over Z}
={1\over Z}\exp\Bigl\{ -\sum_{j=1}^M \beta_j \left(H_j - \mu_j
N_j\right)+ \int_{-\infty}^0 ds e^{\epsilon s}J_S(s)
\Bigr\} , \label{MacLennanZubarevDM}
\end{equation}
where $Z$ is the normalization constant, $1/\beta_j$ and $\mu_j$ are, 
respectively, the
temperature and chemical potential of the $j$th subsystem, 
${\widetilde H}_j\equiv H_j+\int_{-\infty}^0 ds e^{\epsilon s}{dH_j(s)
\over ds}$,
\break
and ${\widetilde N}_j\equiv N_j+\int_{-\infty}^0 ds e^{\epsilon s}
{dN_j(s)\over ds}$
are Zubarev's local integrals of motion\cite{Zubarev}, and\break
$H_j(s)=e^{iHs/\hbar}H_je^{-iHs/\hbar}$, 
$N_j(s)=e^{iHs/\hbar}N_je^{-iHs/\hbar}$ with $H\equiv \sum_jH_j+W$
the total Hamiltonian. 
The integrand in the left-hand side is given by
$J_S(s)= \sum_{j=1}^M \beta_j J_j^q(s)$
where $J_j^q(s)\equiv {dH_j(s)\over ds}- \mu_j {dN_j(s)\over ds}$
stands for non-systematic energy flow, or heat flow, to the $j$th 
subsystem. 
A convergence factor $e^{\epsilon s}$ $(\epsilon >0)$ is introduced
in the time integral, where the limit $\epsilon\to 0$ is taken after 
all the calculations. 
As discussed in Ref.~\citen{Zubarev}, this ensemble well describes 
nonequilibrium phenomena, but it has a 
fundamental difficulty. Indeed, because $J_S(s)$ is the sum of heat 
flows divided by subsystem tempertures, it is the entropy production 
rate of the whole system. 
Hence, if the ensemble (\ref{MacLennanZubarevDM}) would describe
a state consistent with the second law of thermodynamics, the average of 
$J_S(s)$ over $\rho_+$ should be a positive constant
and, as a cosequence, the integral in (\ref{MacLennanZubarevDM}) would 
diverge in the limit of $\epsilon \to 0$.

On the other hand, rigorous researches have been carried out on 
nonequilibrium 
steady states (NESS) of infinitely extended systems and are
developed further in recent years. 
Those include studies on NESSs of harmonic crystals~\cite{SpohnLeb1,Bafaluy}, 
a one-dimensional gas~\cite{Farmer}, unharmonic chains~\cite{Eckmann}, an 
isotropic XY-chain~\cite{HoAraki,XYPillet}, systems with asymptotic 
abelianness~\cite{Ruelle1}, 
a one-dimensional quantum conductor~\cite{ST}, an interacting fermion-spin 
system~\cite{JaksicPillet2}, fermionic junction systems~\cite{Frolich},
a quasi-spin model of superconductors~\cite{Verbeure}, 
a Bose-Einstein condensate in a junction system~\cite{STBose},
a Bose-Einstein condenstate in a small system coupled with
a large reservoir~\cite{Merkli},
a quantum dot coupled with several reservoirs\cite{JaksicPillet4},
on nonequilibrium entropy 
productions~\cite{Ojima1,Ojima2,JaksicPillet1,RuelleEnPro,FrolichEQ} 
and on linear 
responses~\cite{Ruelle1,Ogata}. See also 
reviews~\cite{JaksicPillet3,STMatsui,JaksicPillet4}.
Moreover, we have shown~\cite{TakahashiTasaki1,TakahashiTasaki2} that
NESS constructed by this method
well explains
experiments
on transports of some mesoscopic systems.

In this article, we illustrate the above-mentioned features in terms of 
a spinless-electron model of a single-level quantum dot interacting 
with two two-dimensional reservoirs 
and show that NESS can be regarded as a MacLennan-Zubarev ensemble
in an appropriate sense.  
In the next section, 
the basic features of the C$^*$-algebraic method are summarized
in a less technical way. 
In Sec.~3, a nonequlibrium steady state (NESS) is 
contsructed starting from a local equilibrium state with the aid of the
scattering theory. 
Since general results were proved by Ruelle~\cite{Ruelle1} and 
by Fr\"ohlich, Merkli, Ueltschi~\cite{Frolich}, here we describe
the construction under restricted but practically useful conditions
and apply it to a model of an Ahoronov-Bohm ring with a quantum 
dot\cite{TakahashiTasaki1}. 
In Sec.~4, we show that NESS is an analog to 
the Zubarev-MacLennan ensemble under slightly different 
conditions from Ref.~\citen{STMatsui}.
As an application of this observation, we study a formal 
relation between response and fluctuation at NESS.
The last section is devoted to concluding remarks. 

Before closing this section, we describe the spinless-electron 
model of a single-level quantum dot interacting with two reservoirs.  
The system is described by creation (annihilation) operators of 
the reservoir electron with wave number $k\in {\bf R}^2$: $a_{kr}^{\dagger } (a_{kr})$ 
($r=L$: for the left reservoir,
$r=R$: for the right reservoir)
and the dot electron: $c^\dagger (c)$, which satisfy the canonical 
anticommutation relations:
$$
[a_{kr},a_{k'r'}^{\dagger}]_+= \delta_{rr'}\delta(k-k'){\bf 1} \ , 
\qquad [c,c^\dag]_+= {\bf 1} \ ,
$$
where $[A,B]_+\equiv AB+BA$ is the anticommutator, $\delta_{rr'}$ the 
Kronecker delta, $\delta(k-k')$ 
the Dirac delta function, ${\bf 1}$ the unit operator and the other 
anticommutators vanish. 
The total Hamiltonian is $H=H_L+H_R+H_D+W$ where 
\begin{eqnarray}
H_{r}\equiv \int dk\omega _{kr}a_{k r}^{\dagger }a_{k r }
\qquad
(r =L \ {\rm or } \ R)
\end{eqnarray}
are the reservoir Hamiltonians, $H_D=\epsilon_0 c^\dagger c$ the dot 
energy and $W$ stands for the interaction among the reservoirs and the dot.
In the above, $\omega _{kr }$ ($r=L,R$) stands for the single-electron 
energy of reservoir electrons:
$\omega_{kL}=\omega_{kR}-eV=\hbar^2 k^2/(2m)-eV/2$ where $V$ is the bias 
voltage,
$e$ the elementary charge, 
$\hbar$ the Planck constant and $m$ the effective mass, and $\epsilon_0$ 
is the 
energy of the dot level. 
The numbers of particles in the reservoirs and the dot are given, 
respectively, by $N_{L/R}=\int dka_{k L/R}^\dagger a_{k L/R}$ 
and $N_D=c^\dag c$.

\section{C$^*$-algebraic approach}
\subsection{C$^*$-algebra and Time Evolution\cite{Bratteli,Haag}}
An essential feature of the C$^*$-algebraic method is to discuss the 
properties of 
infinitely extended systems through the investigation of finite 
observables. 
We start from a set ${\cal F}$ of operators $A$ such that the maximum 
eigenvalue (more precisely, 
the maximum spectrum) of $A^\dag A$ is finite and its square root, 
denoted as $\Vert A\Vert$, 
is used for measuring the size of $A$ (or $\Vert\cdot\Vert$ is a norm). 
The set ${\cal F}$ is a complex linear space where the product and the 
`conjugation' $A\to A^\dag$ are 
defined\footnote{In the mathematical literatures such as Ref.~\citen{Bratteli}, 
$A^*$ is used instead of $A^\dag$.},
and the norm $\Vert\cdot\Vert$ satisfies 
(i) \ $\Vert A\Vert\ge 0$ and $\Vert A\Vert= 0$ implies $A=0$, 
(ii) \ $\Vert \alpha A+B\Vert \le |\alpha| \Vert A\Vert+\Vert B\Vert$ 
($\alpha\in {\bf C}, A,B\in {\cal F}$), 
(iii) \ $\Vert AB\Vert\le\Vert A\Vert \Vert B\Vert$ and 
(iv) \ the C$^*$-property: $\Vert A^\dag A\Vert=\Vert A\Vert^2$.
Also ${\cal F}$ is complete with respect to this
norm\footnote{Namely, if $\{A_n\}_{n=1}^\infty \subset {\cal F}$ is 
a sequence such that $\Vert A_n-A_m\Vert \to 0$
for $n,m\to \infty$, then there
is $A\in {\cal F}$ such that $\Vert A_n-A\Vert \to 0$
as $n\to \infty$, or $\{A_n\}_{n=1}^\infty$ has a limit in ${\cal F}$.}.
Such ${\cal F}$ is called a C$^*$-algebra\cite{Bratteli,Haag}.

For the spinless electron model of a quantum dot, ${\cal F}$ is a set of 
operators which can be approximated,
with arbitrary precision, by a finite 
sum\footnote{Let $A\in {\cal F}$, then, for arbitary $\epsilon>0$, 
there exist $\alpha$, $C_\zeta$ and 
$b_j(\zeta)$ ($j=1,\cdots N_\zeta$) 
such that 
$\Vert A-\{\alpha {\bf 1}+\sum_{\zeta} C_{\zeta} b_1(\zeta)^\dag 
\cdots b_s(\zeta)^\dag b_{s+1}(\zeta) 
\cdots b_{N_\zeta}(\zeta)\}\Vert<\epsilon$
holds.}: 
\begin{equation}
\alpha {\bf 1}+\sum_{\zeta} C_{\zeta} b_1(\zeta)^\dag \cdots 
b_s(\zeta)^\dag b_{s+1}(\zeta) \cdots b_{N_\zeta}(\zeta)
\ , \label{finiteSum}
\end{equation}
where $\alpha$, $C_{\zeta}$ are complex numbers, ${\bf 1}\in {\cal F}$ 
is the unit 
and $b_j(\zeta)$ 
($j=1,\cdots N_\zeta$) is either $c$ or
$a_r(f)\equiv \int dk f(k)^* a_{kr}$ ($r=L,R$) with $f(k)$ a square 
integrable function over ${\bf R}^2$,
i.e., $f\in {\rm L}^2({\bf R}^2)$. 
Namely, the algebra ${\cal F}$ is generated by ${\bf 1}, c, a_L(f)$ and 
$a_R(f)$.
Because of the canonical anticommutation relations, one has 
$[a_r(f),a_{r'}(g)]_+=0$, 
$[a_r(f),a_{r'}(g)^\dag]_+=(f,g) \delta_{rr'} {\bf 1}$
where $(f,g)=\int dk f(k)^* g(k)$.
The definition of ${\cal F}$ is meaningful since $a_r(f)$ is bounded.
Indeed, the C$^*$-property leads to
\begin{eqnarray}
&&\Vert a_r(f)\Vert^4 = \Vert a_r(f)^\dag a_r(f)\Vert^2 = \Vert a_r(f)^\dag a_r(f)a_r(f)^\dag a_r(f)\Vert \cr
&&= \Vert a_r(f)^\dag \{(f,f)-a_r(f)^\dag a_r(f)\} a_r(f)\Vert = (f,f) \Vert a_r(f)\Vert^2
\nonumber
\end{eqnarray}
or $\Vert a_r(f)\Vert =\sqrt{(f,f)}<+\infty$.
For later use, we introduce a subset ${\cal F}_{\rm res} (\subset {\cal F})$ 
of reservoir operators, each element of which is approximated, 
with arbitrary precision, by a finite sum: $\alpha {\bf 1}+
\sum_{\zeta} C_{\zeta} a_{r_1}(f_{\zeta,1})^\dag 
\cdots a_{r_s}(f_{\zeta,s})^\dag a_{r_{s+1}}(f_{\zeta,s+1}) \cdots 
a_{r_{N_\zeta}}(f_{\zeta,N_\zeta})$ ($r_j=L$ or $R$, $f_{\zeta,j}\in {\rm L}^2({\bf R}^2)$,
and $\alpha, C_\zeta$ are complex numbers).

Now we turn to the description of the time evolution. 
Because of their unboundedness, 
the Hamiltonians $H_L$, $H_R$ are not included in ${\cal F}$ 
and, thus, are not observables. 
But, the time evolution can be defined within the framework of ${\cal F}$. As an example,
let us consider the free evolution $e^{iH_0t/\hbar}Ae^{-iH_0t/\hbar}$ 
generated by\break
$H_0\equiv~H_L+H_R+H_D$.
A formal calculation gives
$e^{iH_0t/\hbar}a_r(f) e^{-iH_0t/\hbar}= \int dk f(k,t)^* a_{kr}$ 
where $f(k,t)\equiv f(k) e^{i\omega_{kr}t/\hbar}\in {\rm L}^2({\bf R}^2)$, and
$e^{iH_0t/\hbar}c e^{-iH_0t/\hbar}=e^{-i\epsilon_0 t/\hbar}c$, both of which are
elements of ${\cal F}$. Thus, one can define the time evolution of the finite sum
(\ref{finiteSum}) and, thus, of any element of ${\cal F}$. 
Hereafter, to avoid the explicit use of $H_0$, the evolution 
is denoted as
$\tau_t^{(0)}(A) \equiv e^{iH_0t/\hbar}A e^{-iH_0t/\hbar}$, then 
the map 
$A\to \tau_t^{(0)}(A)$
is linear and preserves the product, conjugation and norm: $\tau_t^{(0)}(AB)=\tau_t^{(0)}(A)\tau_t^{(0)}(B)$,
$\tau_t^{(0)}(A)^\dag=\tau_t^{(0)}(A^\dag)$ and $\Vert\tau_t^{(0)}(A)\Vert=\Vert A\Vert$.
Its infinitesimal generator is given by 
$\hat\delta(A)\equiv \left.{d\over dt}\tau_t^{(0)}(A)\right|_{t=0}$ 
on some dense subset $D(\hat\delta)\subset {\cal F}$, called the domain 
of $\hat\delta$\footnote{Namely, any $A\in {\cal F}$ can be approximated by an 
element of $D(\hat\delta)$
with arbitrary precision.}.
For instance, 
$\hat\delta(a_r(f))=-i/\hbar \int dk \omega_{kr} f(k)^* a_{kr}$ ($r=L,R$)
is meaningful only when $\omega_{kr} f(k)\in {\rm L}^2({\bf R}^2)$, 
and a set of finite sums generated by such $a_r(f)$ ($r=L,R$) together with
$\bf 1$ and $c$ provide the domain $D(\hat\delta)$ of $\hat\delta$. 
The evolution $\tau_t$ generated by the total Hamiltonian $H=H_0+W$ is 
defined as a solution of
$$
{d\over dt}\tau_t(A)=\tau_t\left(\hat\delta(A)+{i\over \hbar}[W,A]\right) \ ,
\quad (^\forall A \in D(\hat\delta)) \ ,
$$
under the initial condition $\tau_t(A)|_{t=0}=A$. The map $\tau_t$ has similar 
properties as $\tau_t^{(0)}$. 

\subsection{States\cite{Bratteli,Haag}}\label{subsec:states}
Usually, a statistical state is given by a density matrix. 
However, 
within the algebraic 
approach, it is specified by listing the average value 
$\langle A\rangle$ of an arbitrary element 
$A\in {\cal F}$, i.e., by a complex-valued linear map,
called a normalized positive linear functional:
$A\to \langle A\rangle$,
which satisfies 
$\langle A^\dag A\rangle\ge 0$, $\langle {\bf 1}\rangle=1$ and 
$|\langle A\rangle|\le \Vert A\Vert$. 

Within the algebraic approach, canonical states are formulated without 
explicit reference to the Hamiltonian.
Remind that the grand canonical state with temperature $\beta^{-1}$ 
and chemical potential $\mu$ of a 
finite degree-of-freedom system is given by 
$\langle A\rangle_{gc}={\rm Tr}(Ae^{-\beta(H-\mu N)})/Z_{gc}$ 
with $H$ the Hamiltonian, $N$ the total number of particles and $Z_{gc}$ 
the normalization constant.
Then, it is easy to see that the Kubo-Martin-Schwinger (KMS) boundary 
condition\cite{Kubo,Martin} 
$\langle A\sigma_{i\beta}^{gc}(B)\rangle_{gc}=\langle BA\rangle_{gc}$
is satisfied with respect to
$\sigma_s^{gc}(A)~=~e^{i(H-\mu N)s}Ae^{-i(H-\mu N)s}$.
For infinite systems, the KMS condition {\it defines} canonical 
ensembles\cite{Bratteli,Haag}. 
As an example, let us
consider the grand canonical state with temperature $\beta^{-1}$ 
and chemical potential $\mu$
of the reservoir system described by ${\cal F}_{\rm res}$.
Let
$\sigma_s^{gc}(a_r(f))~\equiv~\int~dkf(k)^*
e^{-i(\omega_{kr}-\mu)s}a_{kr}$\break
($r=L, R$) and ${\cal F}_{\rm res}^{a,gc}$ be a dense set
such that, for any $A\in{\cal F}_{\rm res}^{a,gc}$,
$\sigma_s^{gc}(A)$ 
is analytic in $|{\rm Im}s|\le \beta$\footnote{
${\cal F}_{\rm res}^{a,gc}$ can be a set
of finite sums generated by $a_r(f)$ with
$f(k) e^{i(\omega_{kr}-\mu)s} \in $L$^2({\bf R}^2)$ 
($|$Im~$s|\le \beta$).}, 
then the grand canonical state is defined 
as a state satisfying
$$
\langle A\sigma_{i\beta}^{gc}(B)\rangle_{gc}=\langle BA\rangle_{gc} \ , 
\quad (A,B\in {\cal F}_{\rm res}^{a,gc}) \ .
$$
This equation and the canonical anticommutation relation lead to
$$
\langle a_r^\dag(f)\{ \sigma_{i\beta}^{gc}(a_{r'}(g))+a_{r'}(g)\}
\rangle_{gc}=
\langle [a_{r'}(g),a_r^\dag(f)]_+\rangle_{gc}=
\delta_{rr'} \ (g,f)=\delta_{rr'} \ \int dk g(k)^* f(k) \ .
$$
Since $\sigma_{i\beta}^{gc}(a_{r'}(g))+a_{r'}(g)
=\int dk g(k)^* (e^{\beta(\omega_{kr'}-\mu)}+1)a_{kr'}$, by 
replacing $g(k)$ by $g(k)F(\omega_{kr}) $ with 
$F(x)\equiv 1/\{e^{\beta(x -\mu)}+1\}$ the Fermi
distribution function, one obtains
$$
\langle a_r^\dag(f)a_{r'}(g)\rangle_{gc} 
= \delta_{rr'} \ \int dk g(k)^* f(k) F(\omega_{kr}) \ .
$$
In the same way, one can show that the state 
$\langle \cdots \rangle_{gc}$ 
satisfies Wick's theorem.
In short, the KMS condition fully determines the state 
$\langle \cdots \rangle_{gc}$. 
In general, if no phase transition takes place, the KMS state is unique
and, if a phase transition occurs,  
several KMS states exist as a result of 
spontaneous symmetry breaking.

A local equilibrium state can also be defined as a KMS state. As an example,
we consider a local equilibrium state
where the left (right) reservoir is in the equilibrium state with 
temperature $\beta_L^{-1}$ ($\beta_R^{-1}$)
and chemical potential $\mu_L$ ($\mu_R$). Consider a map 
$\sigma_s$ formally expressed as $\sigma_s(A)=e^{i\sum_r\beta_r(H_r-\mu_r N_r)s} A
e^{-i\sum_r\beta_r(H_r-\mu_r N_r)s}$ 
and defined by 
$\sigma_s(a_r(f)) \equiv \int dk 
f(k)^* e^{-i\beta_r(\omega_{kr}-\mu_r)s}a_{kr}$ ($r=L,R$)
and $\sigma_s(c)=c$, 
then a local equilibrium state
$\langle \cdots \rangle_{loc}$ is given by a KMS condition 
$\langle A\sigma_{i}(B)\rangle_{loc}=\langle BA\rangle_{loc}$
($A\in {\cal F}, B\in{\cal F}_{\rm res}^a$), 
where $B\in {\cal F}_{\rm res}^a$ implies that
$\sigma_s(B)$ is analytic in $|{\rm Im}s|\le 1$.
It again satisfies Wick's theorem and its nonvanishing two-point functions
are
\begin{eqnarray}
\langle a_r^\dag(f)a_r(g)\rangle_{loc} = \int dk g(k)^* f(k) F_r(\omega_{kr}) \  \ (r=L,R) \ ,
\quad \langle c^\dag c\rangle_{loc}={1\over 2}
\end{eqnarray}
where $F_r(x)=1/\{e^{\beta_r(x-\mu_r)}+1\}$ ($r=L,R$) are the Fermi distribution functions.

\subsection{Ergodicity\cite{Bratteli,Haag}}
If the decay of dynamical correlations is sufficiently fast, certain 
states have ergodicity. One of such
dynamical conditions
is the asymptotic abelian property\cite{Bratteli}:
\begin{eqnarray}
\lim_{|t|\to\infty}
\Vert [A,\tau_t(B)]\Vert&=&0 \ , \ ( A \ {\rm or} \ B \in {\cal F}_{\rm even}) \\
\lim_{|t|\to\infty}
\Vert [A,\tau_t(B)]_+\Vert&=&0 \ , \ (A \ {\rm and} \ B \in {\cal F}_{\rm odd})
\end{eqnarray}
where ${\cal F}_{\rm even/odd}\equiv \{A\in {\cal F}: A$ consists of even/odd 
numbers of Fermion operators.$\}$ and
$[A,B]$ stands for the commutator: $[A,B]=AB-BA$.
Then, as stated below (Example 4.3.24 of Ref.~\citen{Bratteli}), 
a clustering property is
satisfied by 
a class of states called `factor' states, which include unique KMS states. 
Roughly speaking, a state $\langle \cdots \rangle$ is called 
`factor'\footnote{
Given a state
$\langle \cdots \rangle$ over a C$^*$-algebra ${\cal F}$, it can be 
represented as a
subalgebra $\pi({\cal F})$ of the algebra ${\cal B}$ of 
all bounded operators on some Hilbert space ${\cal H}$ as
$\langle A \rangle =(\Omega,\pi(A)\Omega)$ with a cyclic 
vector $\Omega\in {\cal H}$
(GNS representation).
Let $\pi({\cal F})^\prime\equiv \{a\in {\cal B}:[a,b]=0, \forall 
b\in
\pi({\cal F})\}$ and $\pi({\cal F})^{\prime\prime}\equiv 
\{c\in {\cal B}:[c,a]=0, \forall a\in
\pi({\cal F})^\prime \}$, then $\langle \cdots \rangle$ is called a 
factor state
iff $\pi({\cal F})^\prime\cap \pi({\cal F})^{\prime\prime}
={\bf C}{\bf 1}$, where ${\bf C}$ is the set of complex numbers and 
${\bf 1}$ is the unit of $\pi({\cal F})$.} 
if any $D\in {\cal F}$ satisfying 
$\langle [A,D] \rangle =0$ ($\forall A\in {\cal F}$) behaves as some 
complex number $\gamma$ in the
sense of $\langle ADB \rangle=\gamma \langle AB \rangle$ 
($\forall A,B \in {\cal F}$).
\medskip

\begin{quote} 
\noindent{\bf Clustering Property}\cite{Bratteli}:
 \ For an asymptotic abelian 
evolution $\tau_t$
and a factor state $\langle \cdots \rangle$, one has
$$
\lim_{|t|\to \infty}\{ \langle A\tau_t(B)C \rangle-\langle AC \rangle 
\langle \tau_t(B) \rangle \}=0 \ .
$$
If $\langle \tau_t(A) \rangle =\langle A\rangle$ ($^\forall A\in {\cal F}$), 
it is mixing:
$\lim_{|t|\to \infty}\langle A\tau_t(B)C \rangle=\langle AC \rangle \langle B\rangle$.
\end{quote}
\medskip

\noindent
For the spinless electron model of a quantum dot,
we show that the local equilibrium state $\langle\cdots\rangle_{loc}$ 
restricted to the subalgebra 
${\cal F}_{\rm res}$ generated by ${\bf 1}, a_L(f)$ and $a_R(f)$ is 
mixing with respect to the evolution $\tau_t^{(0)}$.
The state $\langle\cdots\rangle_{loc}$ is a factor state as a unique 
KMS state and is easily shown to be 
$\tau_t^{(0)}$-invariant. 
On the other hand, for $A,B\in {\cal F}_{\rm res}$, their commutator
$[A,\tau_t^{(0)}(B)]$
can be approximated with arbitrary precision
by a finite sum of the terms like 
$C_t[a_r(f),\tau_t^{(0)}(a_r(g)^\dag)]_+ D_t$ \ 
($C_t,D_t\in {\cal F}_{\rm res}$) and its conjugate.
Then, the Riemann-Lebesgue theorem\cite{FourierTrans} gives
$$
\Vert C_t[a_r(f),\tau_t^{(0)}(a_r(g)^\dag)]_+ D_t\Vert \le 
\sup_t[\Vert C_t\Vert \Vert D_t\Vert]
\left|\int dk f(k)^* g(k) e^{i\omega_{kr}t/\hbar}\right|\to 0
\ \ (|t|\to \infty) \ .
$$
Thus, one has $\lim_{|t|\to \infty}\Vert[A,\tau_t^{(0)}(B)]\Vert=0$, 
or $\tau_t^{(0)}$ is asymptotic abelian. Therefore, as a result of 
Clustering Property, local equilibrium state 
$\langle\cdots\rangle_{loc}$ 
restricted to ${\cal F}_{\rm res}$ is mixing with respect to $\tau_t^{(0)}$:
\begin{equation}
\lim_{|t|\to \infty} \langle A \tau_t^{(0)}(B)C \rangle_{loc}= \langle AC\rangle_{loc} \langle B\rangle_{loc} \ .
\quad (A,B,C\in {\cal F}_{\rm res})
\label{FreeMix}
\end{equation}
Note that the state $\langle\cdots\rangle_{loc}$ is not ergodic on the whole algebra ${\cal F}$ with respect to
$\tau_t^{(0)}$ because $\langle c^\dag \tau_t^{(0)}(c)\rangle_{loc} =e^{-i\epsilon_0t/\hbar}
\langle c^\dag c\rangle_{loc}$ does not converge for $|t|\to \infty$. 

\section{Nonequilibrium Steady States}\label{sematch}
\subsection{Scattering Problem and Nonequilibrium Steady States}
As discussed in Refs.~\citen{SpohnLeb1,Eckmann,HoAraki,Ruelle1,ST,JaksicPillet2,Frolich,Verbeure,STBose,
JaksicPillet1,RuelleEnPro,JaksicPillet3,STMatsui,JaksicPillet4},
a nonequilibrium steady state (NESS) $\langle\cdots\rangle_\pm$ is 
constructed dynamically as an asymptotic
state starting from the local equilibrium state 
$\langle \cdots \rangle_{loc}$  
introduced in Sec.~\ref{subsec:states}:
$
\langle A\rangle_\pm \equiv \lim_{t\to\pm \infty} 
\langle\tau_t(A)\rangle_{loc}
$.
As discussed by Ruelle\cite{Ruelle1} for systems with L$^1$-asymptotic abelian property and by Fr\"ohlich, 
Merkli and Ueltschi\cite{Frolich} for quantum junction systems, the construction of NESS is closely related to 
the scattering problem.
Here we give a restrictive but practically useful characterization used in Refs.\citen{TakahashiTasaki1,TakahashiTasaki2}.

For the spinless model of a quantum dot, 
the interaction $W$ induces scattering of
the left/right-reservoir electrons and the process is 
described by asymptotic fields\cite{Haag}:
\begin{equation}
a_r^{({\rm in/out})}(f) = \lim_{t\to -\infty/+\infty}  \tau_t\left(\tau_t^{(0)-1}(a_r(f))\right) \ , \quad (r=L,R)
\label{ASfield0}
\end{equation}
where $a_r^{({\rm in/out})}(f)$ are incoming/outgoing fields and the limit 
is taken in an approriate sense.
Here we consider a case where the limit (\ref{ASfield0}) exists 
in norm:
\begin{equation}
\lim_{t\to -\infty/+\infty}  \Big\Vert\tau_t\bigl(\tau_t^{(0)-1}(a_r(f))\bigr)-a_r^{({\rm in/out})}(f)\Big\Vert=0 \ , \quad (r=L,R)
\label{ASfield1}
\end{equation}
and the initial state $\langle \cdots \rangle_{loc}$ is $\tau_t^{(0)}$-invariant.
Then, one has\cite{TakahashiTasaki1,TakahashiTasaki2}
\medskip

\begin{quote}
\noindent{\bf Proposition 1}: If, for some subset $D_0\subset {\rm L}^2({\bf R}^2)$,  
(i) the limits (\ref{ASfield1}) exist for any $f\in D_0$, 
(ii) the fields $ a_r^{({\rm in/out})}(f)$ ($r=L, R$, $f\in D_0\subset {\rm L}^2({\bf R}^2)$)
generate the whole algebra ${\cal F}$ and  (iii)
$\langle \cdots \rangle_{loc}$ is $\tau_t^{(0)}$-invariant, then, the limit
\begin{equation}
\langle A\rangle_{+/-} \equiv \lim_{t\to+\infty/-\infty} \langle\tau_t(A)\rangle_{loc} \ ,
\quad (^\forall A\in {\cal F})
\end{equation}
exists and defines a (nonequilibrium) $\tau_t$-invariant state $\langle\cdots\rangle_{+/-}$. Moreover, 
\begin{eqnarray}
&&\langle a_{r_1}^{({\rm out})}(f_1)^\dag \cdots a_{r_s}^{({\rm out})}(f_s)^\dag a_{r_{s+1}}^{({\rm out})}(f_{s+1})
\cdots a_{r_N}^{({\rm out})}(f_N) \rangle_- 
\cr
&&~=\langle a_{r_1}^{({\rm in})}(f_1)^\dag \cdots a_{r_s}^{({\rm in})}(f_s)^\dag a_{r_{s+1}}^{({\rm in})}(f_{s+1})
\cdots a_{r_N}^{({\rm in})}(f_N) \rangle_+ 
\cr
&&~=\langle a_{r_1}(f_1)^\dag \cdots a_{r_s}(f_s)^\dag a_{r_{s+1}}(f_{s+1})
\cdots a_{r_N}(f_N) \rangle_{loc} \ .
\label{chara1}
\end{eqnarray}
\end{quote}
\medskip

\noindent
{\bf N.B.} \
For a model of a single-level quantum dot coupled with free reservoirs (SEBB model) where
the interaction is bilinear with respect to field operators,
Aschbacher, Jak\v si\'c, Pautrat and Pillet\cite{JaksicPillet4} derive an equivalent characterization to 
Proposition 1, but the present form is applicable, in principle, even to the case where the intraction is not
bilinear.

In the previous section, we have seen that the unperturbed evolution 
restricted to the reservoir
algebra ${\cal F}_{\rm res}$ is mixing. 
As a consequence, as first shown by Ruelle\cite{Ruelle1} (see also Ref.\citen{JaksicPillet4}), the steady 
states $\langle\cdots\rangle_\pm$ are mixing
with respect to the full evolution $\tau_t$ in {\it both} directions
of time:
\medskip

\begin{quote}
\noindent{\bf Proposition 2}: Under the same conditions as Proposition 1,
the steady state $\langle\cdots\rangle_{+/-}$ satisfies
\begin{equation}
\lim_{|t|\to \infty} \langle A\tau_t(B)C\rangle_{+/-}=
\langle AC\rangle_{+/-} \langle B\rangle_{+/-} \ , (A,B,C\in {\cal F}) .
\end{equation}
\end{quote}
\medskip

\noindent{\it Proof of Proposition 1}: 
It can be shown immediately from the following lemma:
\medskip

\begin{quote}
\noindent{\bf Lemma 3}: \ If the limits (\ref{ASfield1}) exist and the 
fields $ a_r^{({\rm in/out})}(f)$ ($r=L, R$)
generate the whole algebra ${\cal F}$,  
there exists $\gamma_{+/-}(A)\in {\cal F}$
for any $A\in {\cal F}$ such that
$\lim_{t\to-\infty/+\infty}\Vert A- 
\tau_t\bigl(\tau_t^{(0)-1}(\gamma_{+/-}(A))\bigr)\Vert=0$.
The map $\gamma_\pm$ is a
*-isomorphism between ${\cal F}$ and 
the reservoir algebra ${\cal F}_{\rm res}$
generated by $a_r(f)$ ($r=L,R$), namely,
(i)  $\gamma_\pm(A)\in {\cal F}_{\rm res}$, 
(ii)  $\gamma_\pm(\alpha A+B)=\alpha \gamma_\pm(A)+\gamma_\pm(B)$,
(iii)  $\gamma_\pm(AB)=\gamma_\pm(A)\gamma_\pm(B)$,  (iv) 
$\gamma_\pm(A^\dag)=\gamma_\pm(A)^\dag$
and (v) $\gamma_\pm$ is one-to-one and onto. Moreover, (vi)
$\Vert A\Vert=\Vert \gamma_\pm(A)\Vert$ and
\begin{eqnarray}
&&\gamma_-\left(a_{r_1}^{({\rm out})}(f_1)^\dag \cdots a_{r_s}^{({\rm out})}(f_s)^\dag a_{r_{s+1}}^{({\rm out})}(f_{s+1})
\cdots a_{r_N}^{({\rm out})}(f_N) \right) 
\cr
&&~=\gamma_+\left(a_{r_1}^{({\rm in})}(f_1)^\dag \cdots a_{r_s}^{({\rm in})}(f_s)^\dag a_{r_{s+1}}^{({\rm in})}(f_{s+1})
\cdots a_{r_N}^{({\rm in})}(f_N) \right) 
\cr
&&~=a_{r_1}(f_1)^\dag \cdots a_{r_s}(f_s)^\dag a_{r_{s+1}}(f_{s+1})
\cdots a_{r_N}(f_N) \ .
\label{chara2}
\end{eqnarray}
The maps $\gamma_\pm$ are nothing but the M\o ller morphisms\cite{Bratteli,Ruelle1,JaksicPillet4,STMatsui}:
\begin{equation} 
\lim_{t\to+\infty/-\infty}\Vert \tau_t^{(0)-1}\bigl(\tau_t(A)\bigr)- 
\gamma_{+/-}(A)\Vert=0 \ . \label{DefGamma}
\end{equation}
\end{quote}
\medskip

\noindent 
Indeed, this lemma and $\tau_t^{(0)}$-invariance of 
$\langle\cdots\rangle_{loc}$ give
\begin{eqnarray}
\Big|
&&\langle \tau_t(A)\rangle_{loc}-\langle \gamma_\pm(A) \rangle_{loc}\Big| 
=\left|
\Big\langle \tau_t\Bigl(A-\tau_{-t}\bigl(\tau_{-t}^{(0)-1}(\gamma_\pm(A))
\bigr)\Bigr) \Big\rangle_{loc}\right|
\cr
&&\le \left\Vert \tau_t\Bigl(A-\tau_{-t}\bigl(\tau_{-t}^{(0)-1}
(\gamma_\pm(A))\bigr)\Bigr) \right\Vert
=\Vert A-\tau_{-t}\bigl(\tau_{-t}^{(0)-1}(\gamma_\pm(A))\bigr)
\Vert
\to 0 \ \ (t\to \pm\infty) \ , \nonumber
\end{eqnarray}
or $\langle A\rangle_\pm \equiv \lim_{t\to\pm\infty}
\langle \tau_t(A)\rangle_{loc}= \langle \gamma_\pm(A) \rangle_{loc}$
exists. Clearly, the map $A\to \langle A\rangle_\pm$ is linear and, as
$\displaystyle \langle A^\dag A\rangle_\pm=
\lim_{t\to\pm\infty}\langle \tau_t(A)^\dag \tau_t(A) \rangle_{loc}\ge 0$
and
$\displaystyle \langle{\bf 1}\rangle_\pm=
\lim_{t\to+\infty}\langle \tau_t({\bf 1})\rangle_{loc}=1$,
$\langle \cdots\rangle_\pm$ defines a state.
And it is invariant:
$
\langle \tau_t(A)\rangle_\pm\equiv \lim_{t'\to\pm\infty}
\langle \tau_{t+t'}(A)\rangle_{loc}=\langle A\rangle_\pm \ .
$\break
By substituting
$A=a_{r_1}^{({\rm in/out})}(f_1)^\dag \cdots 
a_{r_s}^{({\rm in/out})}(f_s)^\dag 
a_{r_{s+1}}^{({\rm in/out})}(f_{s+1})
\cdots a_{r_N}^{({\rm in/out})}(f_N)$ into 
$\langle A\rangle_\pm=\langle \gamma_\pm(A)\rangle_{loc}$,
(\ref{chara1}) immediately follows from (\ref{chara2}).
 \ ({\it Q.E.D.}) 

\medskip

\noindent{\it Proof of Proposition 2}: We only show it in case
of $\langle\cdots\rangle_+$.
 Eq.(\ref{DefGamma}) gives
\begin{eqnarray} 
&&\Vert \tau_t^{(0)}(\gamma_+(A))- 
\gamma_+(\tau_t(A))\Vert
\le
\Vert \tau_t^{(0)}(\gamma_+(A))- \tau_t^{(0)}\bigl(\tau_s^{(0)-1}
(\tau_s(A))\bigr)\Vert \cr
&&\mskip 120mu +
\Vert \tau_{s-t}^{(0)-1}\bigl(\tau_{s-t}\bigl(
\tau_t(A)\bigr)\bigr)- 
\gamma_+(\tau_t(A))\Vert \to 0 \ (s\to +\infty) \ ,
\nonumber
\end{eqnarray}
i.e., $\tau_t^{(0)}(\gamma_+(A))=\gamma_+(\tau_t(A))$ ($^\forall t\in 
{\bf R}$).
On the other hand, as shown in the proof of 
Proposition 1, we have $\langle A\rangle_+=\langle \gamma_+(A)
\rangle_{loc}$ and $\gamma_+(A)\in {\cal F}_{\rm res}$, thus,
(\ref{FreeMix}) gives
\begin{eqnarray}
&&\lim_{|t|\to \infty} \langle A\tau_t(B)C\rangle_+=
\lim_{|t|\to \infty} \langle \gamma_+(A)\gamma_+(\tau_t(B))
\gamma_+(C)\rangle_{loc} 
\cr
&&~=\lim_{|t|\to \infty} \langle \gamma_+(A)\tau_t^{(0)}(\gamma_+(B))
\gamma_+(C)\rangle_{loc} 
=\langle \gamma_+(A)\gamma_+(C)\rangle_{loc} 
\langle \gamma_+(B)\rangle_{loc}
\cr
&&~=
\langle AC\rangle_+ \langle B\rangle_+ \ .
\qquad \mskip 300 mu (Q.E.D.) \nonumber
\end{eqnarray}

\medskip

\noindent{\it Proof of Lemma 3}: 
Let us consider the case of incoming fields. Because of (\ref{ASfield1}) 
and
$\big\Vert\tau_t\bigl(\tau_t^{(0)-1}(a_r(f)^\dag\bigr)\big\Vert= 
\big\Vert a_r(f)^\dag\big\Vert$,
one has 
\begin{eqnarray}
&&\big\Vert\tau_t\bigl(\tau_t^{(0)-1}(a_r(f)^\dag a_{r'}(g) )\bigr)- a_r^{({\rm in})}(f)^\dag
a_{r'}^{({\rm in})}(g)\big\Vert 
\le 
\big\Vert a_r(f)^\dag\big\Vert
\big\Vert\tau_t\bigl(\tau_t^{(0)-1}(a_{r'}(g))\bigr) - a_{r'}^{({\rm in})}(g)\big\Vert \cr
&&\mskip 200mu +\big\Vert\tau_t\bigl(\tau_t^{(0)-1}(a_r(f)^\dag)\bigr)- a_r^{({\rm in})}(f)^\dag
\big\Vert
\big\Vert a_{r'}^{({\rm in})}(g)\big\Vert \to 0 \ \ (t\to -\infty) \ .
\nonumber
\end{eqnarray}
Repeating the same arguments, one finds 
\begin{equation}
\lim_{t\to-\infty}\Vert
\tau_t\bigl(\tau_t^{(0)-1}(a_{r_1}(f_1)^{\natural_1} \cdots a_{r_N}(f_N)^{\natural_N})\bigr)-
a_{r_1}^{({\rm in})}(f_1)^{\natural_1} \cdots a_{r_N}^{({\rm in})}(f_N)^{\natural_N}\Vert=0 \ ,
\label{Natu}
\end{equation}
where $\natural_j$ ($j=1,\cdots N$) stands for $\dag$ or no symbol. Therefore, for any finite
sum:
\begin{equation}
A_f\equiv \alpha {\bf 1}+\sum_{\zeta} C_{\zeta} a_{r_1}^{({\rm in})}(f_{\zeta,1})^\dag 
\cdots a_{r_s}^{({\rm in})}(f_{\zeta,s})^\dag a_{r_{s+1}}^{({\rm in})}(f_{\zeta,s+1}) \cdots 
a_{r_{N_\zeta}}^{({\rm in})}(f_{\zeta,N_\zeta})
\ , \label{Fsum}
\end{equation}
where $\alpha,C_\zeta\in{\bf C}$, there exists
\begin{equation}
B_f\equiv \alpha {\bf 1}+\sum_{\zeta} C_{\zeta} a_{r_1}(f_{\zeta,1})^\dag 
\cdots a_{r_s}(f_{\zeta,s})^\dag a_{r_{s+1}}(f_{\zeta,s+1}) \cdots 
a_{r_{N_\zeta}}(f_{\zeta,N_\zeta})\in {\cal F}_{\rm res}
\label{FsumR}
\end{equation}
such that $\lim_{t\to-\infty}\Vert A_f- \tau_t\bigl(\tau_t^{(0)-1}(B_f)
\bigr)\Vert=0$.
On the other hand, as the fields $a_r^{({\rm in})}(f)$ ($r=L,R$) 
generate ${\cal F}$, any $A\in {\cal F}$ can be
approximated by a finite sum (\ref{Fsum}) with arbitrary precision. 
Thus, for any $A\in {\cal F}$, 
there exists $B\in {\cal F}_{\rm res}$ such that 
$\displaystyle\lim_{t\to-\infty}\Vert A- \tau_t\bigl(\tau_t^{(0)-1}(B)\bigr)\Vert=0$.
As such $B$ is unique, we define $B\equiv\gamma_+(A)$. Then, the properties (ii), (iii), (iv) 
and (vi) can be shown immediately. For example, (vi) follows from
$\Big|\Vert \gamma_+(A)\Vert -\Vert A\Vert\Big| 
\le \Vert \tau_t\bigl(\tau_t^{(0)-1}(\gamma_+(A)\bigr)-A\Vert \to 0 \ \ (t\to -\infty)$.
The property (vi) implies that 
$\gamma_+$ is one-to-one. 
Moreover, since any $B\in {\cal F}_{\rm res}$ can be approximated by a finite sum (\ref{FsumR}), one can find
$A\in {\cal F}$ such that $\gamma_+(A)=B$ or $\gamma_+$ is onto.
The second equality of (\ref{chara2}) follows from the definition of $\gamma_+$ and
(\ref{Natu}).
The properties of $\gamma_-$ can be proved in the same way. 
\ \ ({\it Q.E.D.})

\subsection{Nonequilibrium Steady States for Ahoronov-Bohm Ring with Quantum Dot}
In this subsection, we further investigate the properties of 
NESS $\langle \cdots\rangle_+$
when the reservoir-dot interaction is described by a sum of two 
tunneling interactions\cite{TakahashiTasaki1}
\begin{equation}
W=\int dk \{u_{kL} a_{k L}^{\dagger }c+u_{kR} a_{k R}^{\dagger }c\}
+
w e^{i\varphi}
\int dk dq \ u_{kL} u_{qR} \ a_{k L}^{\dagger} a_{q R} +({\rm h.c.})
\ , \label{original H}
\end{equation}
where the first term corresponds to a tunneling via a quantum dot and
the second to a direct tunneling between the two reservoirs.
Real parameters $w$ and $\varphi$ are, respectively, 
the relative strength and phase between the two processes.
This model describes an Ahoronov-Bohm (AB) ring
with a quantum dot\cite{TakahashiTasaki1} and, when $w=0$, it reduces 
to a model of a single-level quantum dot embedded between two reservoirs 
studied in 
Ref.~\citen{JaksicPillet4}.
The tunneling 
matrix elements are assumed to satisfy:
\begin{itemize}

\item[(a)] The real-valued functions $u_{kr}$ ($r=L,R$) are infinitely 
differentiable with respect to $k\in {\bf R}^2$
and $u_{kr}=0$ when $|k|\le k_0$ or $|k|\ge k_1$ (for some $0<k_0<k_1$). 
Then, the functions 
$
\Gamma_r(\omega)\equiv
2\pi \int dk |u_{kr}|^2 \delta(\omega- \omega_{kr} ) 
$ ($r=L,R$)
are integrable and infinitely differentiable on the whole real axis.

\item[(b)] Let $M_r(z)\equiv \int dk |u_{kr}|^2/(z-\omega_{kr})$, then 
the function
$$
\mskip -25mu
\Lambda(z)=\big(1-w^2 M_L(z)M_R(z)\big)(z-\epsilon_0)-\sum_{r=L,R}M_r(z)-
2w\cos\varphi M_L(z)M_R(z) 
$$
has no real zeros and, hence,
$\displaystyle 1/\Lambda_\pm(\omega)\equiv \lim_{\epsilon>0 ,
\epsilon \to 0}{1/\Lambda(\omega\pm i\epsilon)}$
is 
bounded.
\end{itemize}

\subsubsection{Construction of NESS}
To derive incoming fields, it is enough to study the evolution 
$\tau_{-t}\tau_t^{(0)}(a_{r_0}(f))$ where $f(k)$ is in a set
C$_0^\infty({\bf R}^2)$ of 
infinitely differentiable localized functions\footnote{More precisely,
functions with compact support.} as
C$_0^\infty({\bf R}^2)$
is dense in L$^2({\bf R}^2)$.
Then,
thanks to the bilinearity of $W$, $\tau_{-t}(a_{r_0}(f))$ is 
written
as
$$
\tau_{-t}(a_{r_0}(f)) =\sum_{r'=L,R} a_{r'}(\psi_{r'}(t)) + \psi_c(t)^* c \ .
$$
From the equation of motion ${d\over dt}\tau_{-t}(a_{r_0}(f))=
-\hat\delta(\tau_{-t}(a_{r_0}(f)))
-{i\over \hbar}[W,\tau_{-t}(a_{r_0}(f))]$,
the functions $\psi_{r'}(k;t)$ and $\psi_c(t)$ are found to satisfy 
\begin{eqnarray}
&&i\hbar{\partial \over \partial t} \psi_{r'}(k;t)=
\omega_{kr'}\psi_{r'}(k;t)+u_{r'}(k)\Big\{\psi_c(t)+we^{i\varphi_{r'}}
\int dk' u_{\bar{r'}}(k') \psi_{\bar{r'}}(k';t) 
\Big\} , 
\cr
&&i\hbar{\partial \over \partial t} \psi_c(t) =\epsilon_0 \psi_c(t)+
\sum_{r'=L,R} \int dk' u_{r'}(k') \psi_{r'}(k';t) \ 
\ , \label{EqMotion}
\end{eqnarray}
with an initial condition: $\psi_{r'}(k;0)=\delta_{r'r_0}f(k)$, 
$\psi_c(0)=0$, where $\bar L=R, \bar R=L$, $\varphi_L=-\varphi_R=\varphi$.
The linear equations (\ref{EqMotion}) can be solved easily 
and one has
\begin{eqnarray}
&&\psi_{r'}(k;t) \! =\! F_{kr'}(t)\!
+ \!\!
\int \! \! dk'\Bigg\{\!{u_{r'}(k)u_{r'}(k')\xi^+_{\bar{r'}}(\omega_{k'r'})
F_{k'r'}(t)
\over \Lambda_+(\omega_{k'\bar{r'}})
(\omega_{k'r'}-\omega_{k r'}+i0)} 
\! +\!
{u_{r'}(k)u_{\bar{r'}}(k')\kappa_{r'}(\omega_{k'\bar{r'}})
F_{k'\bar{r'}}(t)
\over \Lambda_+(\omega_{k'\bar{r'}})
(\omega_{k'\bar{r'}}-\omega_{k r'}+i0)} \!\Bigg\}
\cr
&&\psi_c(t)= 
\sum_{r'=L,R} \int dk {u_{r'}(k)\chi_{\bar {r'}}^+(\omega_{kr'})
\over \Lambda_+(\omega_{kr'})} F_{kr'}(t)
\label{Sol1}
\end{eqnarray}
where $\Lambda_\pm(\omega)$ is introduced in (b),
$\xi_r^\pm(x)\equiv 1+M_r^\pm(x)
\big(2w\cos\varphi+w^2(x-\epsilon_0)\big)$,
$\kappa_r(x)\equiv 1+we^{i\varphi_r}(x-\epsilon_0)$,
$\chi^\pm_r(x)\equiv 1+we^{i\varphi_r}M^\pm_r(x)$
with $\displaystyle 
M_r^\pm(x)=\lim_{\epsilon>0, \epsilon \to 0}M_r(x\pm i\epsilon)$,
$$
F_{kr}(t) \! =\! 
e^{-i{\omega_{kr}t\over \hbar}}
\Bigg[
\psi_r(k)\!
+ \!\!
{u_r(k)\over \Lambda_-(\omega_{kr})}
\int \! \! 
dk'\Bigg\{\!{\xi^-_{\bar r}(\omega_{kr})u_r(k') 
\psi_r(k')
\over 
\omega_{kr}-\omega_{k'r}-i0} 
\! +\!
{\kappa_r(\omega_{kr}) u_{\bar r}(k')
\psi_{\bar r}(k')
\over 
\omega_{kr}-\omega_{k'\bar r}-i0} \!\Bigg\}
\Bigg] \ ,
$$
with $\psi_r(k)=\delta_{r'r_0}f(k)$.
When $f(k)\in$ C$_0^\infty({\bf R}^2)$ and 
the conditions (a), (b) are satisfied, 
the limits like $\displaystyle
\int dk \mskip 3mu h(k)/(x-\omega_{kr}\pm i0)\equiv \lim_{\epsilon\to 0, 
\epsilon>0} \int dk \mskip 3mu h(k)/(x-\omega_{kr}\pm i\epsilon)$\break
converge pointwise and uniformly with respect to $x$.
As a consequence, $\tau_{-t}\tau_t^{(0)}(a_{r_0}(f))$ is again 
a linear combination
\begin{equation}
\tau_{-t}\tau_t^{(0)}(a_{r_0}(f)) =
\sum_{r'=L,R} a_{r'}(\psi^{({\rm in})}_{r_0r'}(f)) 
+ \psi^{({\rm in})}_{r_0c}(f)^* c 
+\sum_{r'=L,R} a_{r'}(\Delta{\widetilde\psi}_{r'}(t)) 
+ \Delta{\widetilde\psi}_c(t)^* c 
\nonumber
\end{equation}
where $\psi^{({\rm in})}_{r_0r'}(k;f)$, $\psi^{({\rm in})}_{r_0c}(f)$ 
are derived from (\ref{Sol1})
by replacing $F_{kr}(t)$ to $\delta_{rr_0}f(k)$:
\begin{eqnarray}
&&\psi^{({\rm in})}_{r_0r_0}(k;f) \equiv f(k)+
\int dk'
{u_{r_0}(k) u_{r_0}(k')^* \xi_{\bar{r_0}}^+(\omega_{k'r_0})
\over \Lambda_+(\omega_{k'r_0})(\omega_{k'r_0}-\omega_{kr_0}+i0)}
f(k') \ ,
\cr
&&\psi^{({\rm in})}_{r_0\bar{r}_0}(k;f) \equiv
\int dk'
{u_{\bar{r}_0}(k) u_{r_0}(k')^* \kappa_{\bar{r}_0}(\omega_{k'r_0})
\over \Lambda_+(\omega_{k'r_0})(\omega_{k'r_0}-\omega_{k\bar{r}_0}+i0)}
f(k') \ ,
\cr
&&\psi^{({\rm in})}_{r_0c}(f) \equiv 
\int dk'
{u_r(k')^* \chi^+_{\bar r_0}(\omega_{k'r_0})\over 
\Lambda_+(\omega_{k'r_0})}f(k') \ ,
\label{Sol3}
\end{eqnarray}
and $\Delta{\widetilde\psi}_{r'}(k;t)$, $\Delta{\widetilde\psi}_c(t)$
are obtained from (\ref{Sol1})
by replacing $F_{kr}(t)$ to
$$
\Delta{\widetilde F}_{kr}(t) \! =\! 
{u_r(k)\over \Lambda_-(\omega_{kr})} \eta_{rr_0}(\omega_{kr})
\int \! \! 
dk'
\!{ u_{r_0}(k') 
f(k')
\over 
\omega_{kr}-\omega_{k'r_0}-i0} 
e^{i(\omega_{k'r_0}-\omega_{kr})t/\hbar}
\ ,
$$
with
$\eta_{r_0r_0}(x)=\xi^-_{\bar r_0}(x)$, $\eta_{\bar r_0r_0}(x)=
\kappa_{\bar r_0}(x)$.
Reminding  
$\displaystyle
\lim_{t\to +\infty}{e^{-ixt}/(x+i0)}=-2\pi i\delta(x)
$
in the sense of distribution, one expects 
$\Delta{\widetilde F}_{kr}(t)\to 0 \ (t\to +\infty)$ and, thus,
\begin{equation}
\tau_{-t}\tau_t^{(0)}(a_{r_0}(f)) \to
\sum_{r'=L,R} a_{r'}(\psi^{({\rm in})}_{r_0r'}(f)) 
+ \psi^{({\rm in})}_{r_0c}(f)^* c 
\equiv a_{r_0}^{({\rm in})}(f)
\ , \ \ (t\to+\infty) .
\label{InField}
\end{equation}
Indeed, if the conditions (a) and (b) are satisfied and 
$f\in$ C$_0^\infty({\bf R}^2)$, 
an argument similar to those of Refs.~\citen{STBose,STAccardi} gives
\begin{eqnarray}
&&\Vert \tau_{-t}\tau_t^{(0)}(a_{r_0}(f))-a_{r_0}^{({\rm in})}(f)\Vert 
\le
\sum_{r'=L,R} \Vert a_{r'}\bigl(\Delta{\widetilde\psi}_{r'}(t)\bigr)\Vert
 + |\Delta{\widetilde\psi}_c(t)| \cr
&&~= \sum_{r'=L,R} \sqrt{\int dk \big|\Delta{\widetilde\psi}_{r'}(k,t)
\big|^2} 
+|\Delta {\widetilde\psi}_c(t)|\to 0 \ , (t\to +\infty) \ .
\nonumber
\end{eqnarray}
Moreover, under the condition (a),
the original operators are expressed as
\begin{equation}
a_r(f)=\sum_{r'=L,R}a_{r'}^{({\rm in})}\bigl(\varphi_{r'r}(f)\bigr)
\ , \qquad
c=\sum_{r'=L,R}a_{r'}^{({\rm in})}\bigl(\varphi_{r'c}\bigr) \ ,
\label{Original1}
\end{equation}
where the functions $\varphi_{r'r}(k;f)$ and $\varphi_{r'c}(k)$ are
given by
\begin{eqnarray}
\varphi_{r'r}(k;f)&=& \delta_{rr'}f(k)+
{u_{r'}(k)\eta_{r'r}(\omega_{kr'})\over \Lambda_-(\omega_{kr'})}
\int dk' {u_r(k') f(k')\over \omega_{kr'}-\omega_{k'r}-i0} \ , \cr
\varphi_{r'c}(k)&=& {u_{r'}(k)\chi^+_{\bar r'}(\omega_{kr'})^*
\over \Lambda_-(\omega_{kr'})} \ .
\label{Original2}
\end{eqnarray}
This implies that the incoming fields generate the whole algebra ${\cal F}$.
Then, because of Proposition 1 and the fact that Wick's theorem
holds for the expectation value of a product of $a_r(f)$ and 
$a_{r'}(f')^\dag$ 
with respect to $\langle\cdots\rangle_{loc}$, we find:
\medskip

\begin{quote}
\noindent{\bf Proposition 4}: \ For the model
where the 
interaction
is given by (\ref{original H}), if the tunneling matrix elements satisfy 
the conditions (a) and (b), then the limit: \
$\lim_{t\to +\infty}\langle \tau_t(A)\rangle_{loc}\equiv \langle A\rangle_+$
exists for any $A\in{\cal F}$ and defines a steady state. Moreover, 
the steady state $\langle\cdots \rangle_+$ satisfies
Wick's
theorem 
with respect to the incoming fields $a_r^{({\rm in})}(f)$ 
introduced in 
(\ref{Sol3}) and (\ref{InField}),
and the nonvanishing two-point functions are given by
\begin{equation}
\langle a_{r_1}^{({\rm in})}(f_1)^\dag 
a_{r_2}^{({\rm in})}(f_2) \rangle_+ =
\langle a_{r_1}^\dag(f_1)a_{r_2}(f_2)\rangle_{loc} = 
\int dk f_2(k)^* f_1(k) F_{r_1}(\omega_{kr_1}) \ ,
\label{NessTwoPoint}
\end{equation}
where 
$F_r(x)=1/\{e^{\beta_r(x-\mu_r)}+1\}$ ($r=L,R$) is the Fermi 
distribution function of inverse temperature $\beta_r$ and
chemical potential $\mu_r$.
\end{quote}
\medskip

\noindent
A simple interpretation could be given to this result. Suppose that there exists
an invariant 
vacuum state $|{\rm vac.}\rangle$,
then, due to the difference between the Schr\"odinger and Heisenberg
pictures, the vector $\tau_{-t}\big(a_{r}^{({\rm in})}(f)^\dag\big)
|{\rm vac.}\rangle$ describes a one-particle state at time $t$ starting
from an initial state: $a_{r}^{({\rm in})}(f)^\dag|{\rm vac.}\rangle$.
When $t\ll 0$, one may regard 
$\tau_{-t}\big(a_{r}^{({\rm in})}(f)^\dag\big){\rm vac.}\rangle~\simeq~\tau_{-t}^{(0)}(a_r(f)^\dag)|{\rm vac.}\rangle$ 
as a consequence of a relation:\break
$\lim_{t\to -\infty}\Vert 
\tau_{-t}^{(0)}(a_r(f)^\dag)-\tau_{-t}(a_r^{({\rm in})}(f)^\dag)
\Vert=0$ (cf. eq.(\ref{ASfield1})).
In this sense, $a_{r}^{({\rm in})}(f)^\dag$ describes
a particle which was an unperturbed particle of the $r$th reservoir in the far past.
Thus, $\langle\cdots\rangle_+$ is a steady state such that
particles carry the temperature 
and chemical potential of the reservoir from which they come. 

\subsubsection{Transports}
As an application of Proposition 4, transports in the steady state 
$\langle\cdots\rangle_+$ will be studied. 
Formally the energy and the particle number of the reservoir are expressed,
respectively, by $H_r=\int dk \omega_{kr}a_{kr}^\dag a_{kr}$ and
$N_r=\int dk a_{kr}^\dag a_{kr}$ ($r=L,R$), and a formal calculation leads to
\begin{eqnarray}
\mskip -40 mu
&&{d\over dt} e^{iHt/\hbar}N_re^{-iHt/\hbar}\Big|_{t=0}=-{i\over\hbar}
\big(a_r(u_r)^\dag \{c+we^{i\varphi_r}a_{\bar r}(u_{\bar r})\} - 
({\rm h.c.})\big) \equiv \! J_r \ ,
\cr
&&{d\over dt} e^{iHt/\hbar}H_re^{-iHt/\hbar}\Big|_{t=0}=-{i\over\hbar}
\big(a_r(u_r^E)^\dag \{c+we^{i\varphi_r}a_{\bar r}(u_{\bar r})\} 
- 
({\rm h.c.})\big) \equiv \! J_r^E \ ,
\end{eqnarray}
where $u_r(k)$ is the tunneling matrix elements, $u_r^E(k)=\omega_{kr}u_r(k)$, 
and $J_r, J_r^E\in {\cal F}$ are defined by the middle expressions.
Therefore, $J_r$ and $J_r^E$ ($r=L,R$) can be regarded as the 
particle and energy flows to the reservoirs.
Then, (\ref{Original1}), (\ref{Original2}) and Proposition~4 give

\medskip

\begin{quote}

\noindent{\bf Proposition 5}: \ The steady state considered in 
Proposition 4 carries
the particle and energy flows:
\begin{eqnarray}
\mskip -30 mu
\langle J_L\rangle_+ &=& -\langle J_R\rangle_+ = {1 \over 2\pi\hbar}
\int_{-\infty}^\infty d\omega T(\omega)
\{F_R(\omega)-F_L(\omega)\} \ , 
\label{MassFl}\\
\mskip -30 mu \langle J_L^E\rangle_+ &=& -\langle J_R^E\rangle_+ = {1 \over 2\pi\hbar}
\int_{-\infty}^\infty d\omega 
T(\omega) \{F_R(\omega)-F_L(\omega)\} \ . 
\label{EnergyFl}
\end{eqnarray}
where
$$
T(\omega)=
\left| {1+we^{i\varphi}(\omega-\epsilon_0)\over \Lambda_+(\omega)}\right|^2
\Gamma_L(\omega) \Gamma_R(\omega) \ .
$$
The entropy production rate $J_S\equiv \sum_r \beta_r J_r^q$, where $J_r^q =J_r^E-\mu_r J_r$
are heat flows to the reservoirs, has the following NESS average
$$
\langle J_S\rangle_+ =
\int_{-\infty}^\infty d\omega 
{T(\omega)\over 2\pi \hbar}
\{F_R(\omega)-F_L(\omega)\}\{
\beta_L(\omega -\mu_L)-\beta_R(\omega -\mu_R)\} \ , \nonumber
$$
which is nonnegative and vanishes if and only if $\beta_L=\beta_R$ and 
$\mu_L=\mu_R$.

\end{quote}

\medskip

\noindent
The nonnegativity of $\langle J_S\rangle_+$ immediately follows from an inequality:
$$
\left({1\over e^y+1}-{1\over e^x+1}\right)(x-y)\ge 0 \ ,
$$
where the equality holds if and only if $x=y$.
As shown in Refs.~\citen{Ojima1,Ojima2,JaksicPillet1,RuelleEnPro},
positivity of the entropy production can be proved 
for more general cases 
since it is 
related to the relative entropy between the initial state and the state
at time $t$: If the two states were described by density matrices, 
respectively, $\rho_{loc}$ and $\rho_t$, the relative entropy 
$
S(\rho_{loc}|\rho_t)\equiv {\rm Tr}\{\rho_t(\log \rho_t-\log \rho_{loc})\}
$
is related to the entropy production via
\begin{equation}
\langle J_S\rangle_t ={d\over dt}S(\rho_{loc}|\rho_t) \ ,
\label{EnPro}
\end{equation}
where $\langle\cdots\rangle_t$ stands for the average with respect to
$\rho_t$. The same relation holds for a C$^*$-generalization of the
relative entropy given by Araki~\cite{Bratteli,ArakiEn,Uhlmann,Petz}.
Then, because of $S(\rho_{loc}|\rho_t)\ge 0$, l'Hospital's
rule gives the positivity of $\langle J_S\rangle_+$\cite{JaksicPillet1}.

For states described by density matrices, 
(\ref{EnPro}) can be easily shown\cite{Ojima1}. 
Then, we have
$\rho_{loc}=e^{-\sum_j\beta_j(H_j-\mu_jN_j)}/Z_0$,
$\rho_t=e^{-iHt/\hbar}\rho_{loc}e^{iHt/\hbar}$
with $Z_0$ a constant, and
$$
S(\rho_{loc}|\rho_t)={\rm Tr}\big\{\rho_{loc}
\big(-\sum_j\beta_j(H_j-\mu_jN_j)+\sum_j\beta_j
(H_j(t)-\mu_jN_j(t))\bigr)\bigr\} \ ,
$$
where 
$H_j(t)=e^{iHt/\hbar}H_je^{-iHt/\hbar}$ and 
$N_j(t)=e^{iHt/\hbar}H_je^{-iHt/\hbar}$,
and which gives (\ref{EnPro}): 
$$
{d\over dt}S(\rho_{loc}|\rho_t)={\rm Tr}\Big\{\rho_{loc}
\sum_j\beta_j
\Big({dH_j(t)\over dt}-\mu_j{dN_j(t)\over dt}\Bigr)\Bigr\}=
\langle J_S(t)\rangle_{loc}=\langle J_S\rangle_t
\ .
$$
This observation indicates that the features consistent with 
thermodynamics come from the second term of the relative entropy: 
$-{\rm Tr}(\rho_t \log \rho_{loc})$,
which is similar to the nonequlibrium entropy of 
Zubarev\cite{Zubarev}: $S_Z=-{\rm Tr}(\rho_t \log \rho_l)$ 
(cf. eq.(22.31) of Ref.~\citen{Zubarev}) with
$\rho_l$ a reference local equilibrium state.
An entropy of Zubarev type (more precisely, the relative 
entropy between the state $\rho_t$ and a reference local equilibrium 
state $\rho_l$: $S(\rho_t|\rho_l)$) was studied by 
Fr\"ohlich, Merkli, Schwarz and Ueltschi\cite{FrolichEQ} in a
slightly different context.

The identification of $J_r^q\equiv J_r^E-\mu_r J_r$ with the heat
flow could be justified since $\int_0^Tdt J_r^q(t)$
behaves as a thermodynamic heat in the weak coupling limit for a small 
system coupled with a single reservoir\cite{STwaseda,STMatSuken}.

Note that the expressions of the particle and energy flows agree with 
those of the 
Landauer formula\cite{LandauerButtiker,Datta}.
As discussed e.g., by Sivan and Imry\cite{Sivan},
when the temperature difference $\beta_L^{-1}-\beta_R^{-1}$ and chemical potential 
differece 
$\mu_L-\mu_R$ are small, (\ref{MassFl}) and (\ref{EnergyFl}) reduce to 
linear relations
between thermodynamic forces and flows, where Onsagar's
reciprocal relations hold. 
The general proofs of linear response relations for nonlocal 
perturbations such as
the temperature difference and/or chemical potential difference are 
discussed by 
Jak\v si\' c, Ogata and Pillet\cite{Ogata}.

\section{KMS Characterization of Nonequilibrium Steady States}
As discussed in Sec.~\ref{subsec:states}, canonical and grand canonical states are characterized as
KMS states. 
As shown for L$^1$-asymptotic abelian systems in Ref.~\citen{STMatsui},
the nonequilibrium steady states discussed in the previous section can be characterized
in a similar way:

\medskip

\begin{quote}
\noindent{\bf Proposition 6}: \ If the limits (\ref{ASfield1}) exist, the fields $ a_r^{({\rm in/out})}(f)$ 
($r=L, R$) generate the whole algebra ${\cal F}$ and $\langle \cdots \rangle_{loc}$ is $\tau_t^{(0)}$-invariant, 
the nonequilibrium steady states $\langle\cdots\rangle_\pm$ are 
KMS states with respect to the maps
\begin{equation}
\sigma_s^{(\pm)} \equiv \gamma_\pm^{-1} \sigma_s \gamma_\pm \ ,
\label{NESS-KMS}
\end{equation}
namely, $\langle A\sigma_i^{(\pm)}(B)\rangle_\pm=\langle BA\rangle_\pm$ for $A,B\in {\cal F}_\pm^a$
where $i=\sqrt{-1}$, $\sigma_x$ is a map defining $\langle\cdots\rangle_{loc}$ as a KMS state, $\gamma_\pm$ are
maps introduced in Lemma~3 and 
${\cal F}_\pm^a \subset {\cal F}$ are (dense) subsets such that 
$\sigma_z^{(\pm)}(A)$ is analytic in $|$Im~$z|\le 1$ for any 
$A\in {\cal F}_\pm^a$.
Note that (\ref{NESS-KMS}) is well-defined as $\sigma_s \gamma_\pm(A)\in {\cal F}_{\rm res}$
($^\forall A\in {\cal F}$).

\hskip 18 pt
Let ${\hat \delta}_\omega(A)\equiv {d\over ds}\sigma_s(A)\big|_{s=0}$
($A\in D({\hat \delta}_\omega)$) and 
${\hat \delta}_\omega^\pm(A)\equiv {d\over ds}\sigma_s^{(\pm)}(A)\big|_{s=0}$ ($A\in D({\hat \delta}_\omega^\pm)$),
where $D({\hat \delta}_\omega)$ and $D({\hat \delta}_\omega^\pm)$ 
are (dense) subsets where the corresponding
derivatives exist. Suppose that $W\in D({\hat \delta}_\omega)$ and 
there exists a dense subset ${\cal G}_\pm$ of ${\cal F}$ such that
\begin{equation}
\int_{{\bf R}^\mp} dt \Vert [ \tau_t\bigl({\hat \delta}_\omega(W)\bigr),A]\Vert <+\infty
\label{L1asymp}
\end{equation}
holds for any $A\in {\cal G}_\pm$, where ${\bf R}^-$ (${\bf R}^+$) stands for the set of non-positive (non-negative)
real numbers, then one has
\begin{equation}
{\hat \delta}_\omega^\pm(A) = {\hat \delta}_\omega(A) + {i\over\hbar} \int_{\mp \infty}^0 dt 
\left[ \tau_t\left({\hat \delta}_\omega(W)\right),A\right] \ , 
\quad (^\forall A\in D({\hat \delta}_\omega)\cap {\cal G}_\pm) \ .
\label{SteadyKMS}
\end{equation}

\end{quote}

\medskip

\noindent
Before going to the proof of Proposition 6, we discuss its implications. 
Remind that 
$\sigma_s(A)=e^{i \sum_{r=L,R}\beta_r(H_r-\mu_r N_r)s}Ae^{-i\sum_{r=L,R}\beta_r(H_r-\mu_r N_r)s}$
for finite-degree-of-freedom systems,
where $H_r$ and $N_r$ ($r=L,R$) are, respectively, the energy and the number of particles in each reservoir.
Hence, ${\hat \delta}_\omega(A)= i\sum_{r=L,R}\beta_r[(H_r-\mu_r N_r),A]$ and
\begin{equation}
{\hat \delta}_\omega(W)= -i \sum_{r=L,R}\beta_r[H,H_r-\mu_r N_r]
=-\sum_{r=L,R}\beta_r{d\tau_t(H_r-\mu_r N_r)\over dt}\Big|_{t=0}=-J_S \ ,
\end{equation}
where $J_S$ is the entropy production operator discussed in Sec.~\ref{Intro}. 
Therefore, if Proposition 6 were
applicable to finite-degree-of-freedom systems, one would have
$$
{\hat \delta}_\omega^+(A) = i\sum_{r=L,R}\beta_r[(H_r-\mu_r N_r),A] + i \int_{-\infty}^0 dt 
\left[ \tau_t(J_S),A\right] \ .
$$
On the other hand, 
if the state is described by a density matrix $\rho\propto e^{-\Gamma}$, 
it satisifes a KMS condition:
$\langle A \sigma_i^\rho(B)\rangle_\rho=\langle BA\rangle_\rho$, where $\sigma_s^\rho(B)=e^{i\Gamma s}Be^{-i\Gamma s}$.
Hence, 
the density matrix of
the steady state $\langle\cdots\rangle_+$, if it exists, 
is 
given by a MacLennan-Zubarev ensemble:
\begin{equation}
\rho_+ 
={1\over Z}\exp\Bigl\{ -\sum_{j=1}^N \beta_j \left(H_j - \mu_j
N_j\right)+ \int_{-\infty}^0 ds J_S(s)
\Bigr\} \ . \label{MacLennanZubarevDM2}
\end{equation}
As pointed out in Sec.~\ref{Intro}, the original proposal 
(\ref{MacLennanZubarevDM2}) by MacLennan and Zubarev
cannot be justified. Rather, KMS states with respect to 
$\sigma_s^{(\pm)}$ 
which is generated by (\ref{SteadyKMS}) 
should be regarded as a precise definition of the 
MacLennan-Zubarev ensembles.

\medskip

\noindent{\it Proof of Proposition 6}: \ When the dynamics is 
L$^1$-asymptotic abelian and M\"oller morphisms are invertible, 
we have shown the same conclusion\cite{STMatsui,STMatSuken}. 
Since the present conditions are different from the previous ones,
we give the outline of the proof in case of $\langle\cdots\rangle_+$.

Remind that $\langle A\sigma_i(B)\rangle_{loc}=\langle BA\rangle_{loc}$
($A,B\in {\cal F}_{\rm res}^a$) holds for some (dense) subset 
${\cal F}_{\rm res}^a\subset {\cal F}_{\rm res}$. 
On the other hand, the map 
$\gamma_+$ defined in Lemma 3 has the inverse $\gamma_+^{-1}$ 
on ${\cal F}_{\rm res}$ and the steady state is given by
$\langle A \rangle_+=\langle \gamma_+(A) \rangle_{loc}$.
Hence, one has 
\begin{equation}
\langle A\gamma_+^{-1}\bigl(\sigma_i(\gamma_+(B))\bigr)\rangle_+
=\langle\gamma_+(A) \sigma_i(\gamma_+(B))\rangle_{loc}
=\langle\gamma_+(B) \gamma_+(A))\rangle_{loc}
=\langle BA\rangle_+ \ ,
\end{equation}
for any two elements of the dense set $\{A| \gamma_+(A)\in 
{\cal F}_{\rm res}^a \}
\subset {\cal F}$.
Namely, the steady state is a KMS state
with respect to $\sigma_s^{(+)}=\gamma_+^{-1}\sigma_s\gamma_+$.
This proves the first half.

Now we consider the generator. Let $\gamma_t(A)\equiv \tau_t^{(0)-1}
\bigl(\tau_t(A)\bigr)$, then, 
for any $A\in {\cal F}$, $\gamma_t(A)$ is differentiable and
$
{d\over dt}\gamma_t(A) = {i\over\hbar} \gamma_t\left([\tau_{-t}(W),A]\right) \ ,
$
which leads to
$$
{d\over dt} \gamma_t^{-1} \sigma_s \gamma_t(A)
={i\over \hbar}
\left[\tau_{-t}\left(\sigma_s(W)-W\right),\gamma_t^{-1} 
\sigma_s \gamma_t(A)\right]
$$
or
\begin{eqnarray}
\gamma_t^{-1} \sigma_s \gamma_t(A)
= \sigma_s(A)+{i\over\hbar} \int_0^t dt' \left[\tau_{-t'}\left(\sigma_s(W)-W\right),
\gamma_{t'}^{-1} 
\sigma_s \gamma_{t'}(A)\right] \ ,
\end{eqnarray}
where we have used $\gamma_t \tau_{-t}=\tau_{-t}^{(0)}$, 
$\tau_{-t}^{(0)}\sigma_s=\sigma_s \tau_{-t}^{(0)}$,
$\gamma_{t}^{-1} \tau_{-t}^{(0)} =\tau_{-t}$. 
Since $W\in D({\hat \delta}_\omega)$ is assumed, for any 
$A\in D({\hat \delta}_\omega)$, one has
\begin{equation}
{\hat \delta}_\omega^{(t)}(A)\equiv \left.{d\over ds}\gamma_t^{-1} \sigma_s 
\gamma_t(A)\right|_{s=0} = {\hat \delta}_\omega(A)+{i\over\hbar} \int_{-t}^0 dt'
\left[\tau_{t'}\left({\hat \delta}_\omega(W)\right),
A \right] \ . \label{GenSug}
\end{equation}
It can be shown that ${\hat \delta}_\omega^{(t)}$ is indeed the 
genrator of
$\{\gamma_t^{-1} \sigma_s \gamma_t\}_{s\in {\bf R}}$\cite{STMatSuken}.
On the other hand, since Lemma 3 implies
\begin{equation}
\lim_{n\to\infty}\Vert\gamma_n^{-1} \sigma_s \gamma_n(A)
- \sigma_s^+(A) \Vert=0 \ ,
\label{UnLim}
\end{equation} 
for any $A\in {\cal F}$, one has
${\hat \delta}_\omega^+(A)=\lim_{n\to \infty}
{\hat \delta}_\omega^{(n)}(A)$, 
if the limit exists\footnote{One can show that the limit (\ref{UnLim})
is uniform on a finite closed $s$-interval. Then, the generator 
${\hat \delta}_\omega^+$ of $\sigma_s^+$ is the graph limit of 
$\{{\hat \delta}_\omega^{(n)}\}_{n\ge 0}$ (cf. Theorem 3.1.28 of
Ref.~\citen{Bratteli}). Also ${\hat \delta}_\omega^{(n)}$
is proved to
have the same domain: $D({\hat \delta}_\omega^{(n)})
=D({\hat \delta}_\omega)$.
Then, if ${\hat \delta}_\omega^{(n)}(A)$
converges, the limit is ${\hat \delta}_\omega^+(A)$\cite{STMatSuken}.
}.
Because of the assumption (\ref{L1asymp}), this is indeed the case for 
$A\in D({\hat \delta}_\omega)\cap {\cal G}_+$ and we obtain 
the desired result (\ref{SteadyKMS}). \hskip 20pt ({\it Q.E.D.})

\medskip
Originally, Kubo\cite{Kubo} used the KMS condition to show the
fluctuation-dissipation relations. Since we have a KMS characterization
of NESS, it is interesting to explore a relation between
response and fluctuation at NESS following Ref.~\citen{Kubo}.

Suppose that 
a spatially uniform electric field of strength $E(t)
=\int_{-\infty}^{+\infty} {d\omega\over 2\pi} 
e^{i\omega t}{\hat E}(\omega)$ is applied to a finite
domain, then,
as shown e.g., 
by Gavish, Imry and Yurke\cite{Gavish},  
its effect is decribed by a
perturbation Hamiltonian: $\int_{-\infty}^t dt' E(t') {\hat I}$
(${\hat I}\in {\cal F}$: spatially averaged
current) and
we have
\begin{equation}
\langle\hat I \rangle_{+,E(t)} 
=\langle\hat I \rangle_+ +
\int_{-\infty}^{+\infty} {d\omega\over 2\pi} 
e^{i\omega t}
G(\omega) {\hat E}(\omega)+{\rm O}(E(t)^2) \ ,
\label{ResponseFunction}
\end{equation}
where 
$\langle\cdots\rangle_{+,E(t)}$ is the state perturbed by
$E(t)$ and  
the
frequency-dependent 
conductance $G(\omega)$ is 
a distribution:
\begin{equation}
G(\omega)={1\over \hbar(\omega-i0)}\int_{-\infty}^0 dt e^{i\omega t} 
\langle[\tau_t({\hat I}),{\hat I}]\rangle_+ \ . \label{NESSconductance}
\end{equation}
When $\hat E(\omega)$ and the Fourier transform of $\langle[\tau_t({\hat I}),{\hat I}]\rangle_+$ 
have appropriate smoothness and integrability,  
the formal calculations can be justified as shown by Ruelle\cite{Ruelle1}.
Note that the zero-frequency limit Re$G(0+)$ of (\ref{NESSconductance}) 
corresponds to a dc-conductance, but is not necessarily agrees with the
differential conductance 
${d\over dV}\langle \hat I\rangle_+$ as shown in 
Ref.~\citen{TakahashiTasaki1} because
the former comes from a local perturbation but the latter from 
a nonlocal perturbation.

Now
we introduce a gauge transformation
$g_\varphi:{\cal F}\to{\cal F}$, which is formally expressed as
$g_\varphi(A)=e^{i(\sum_rN_r+c^\dag c)\varphi}Ae^{-i(\sum_rN_r+c^\dag c)\varphi}$.
Clearly, $g_\varphi$ is a linear map which preserves product and conjugation, and
its action to the generators is $g_\varphi(c)=e^{-i\varphi}c$, 
$g_\varphi(a_r(f))=e^{-i\varphi}a_r(f)$
($r=L,R$). 
Then, we observe that, if the interaction $W$ is gauge-invariant: $g_\varphi(W)=W$,
the maps $\tau_t$, $\sigma^{(+)}_s$ (cf. Proposition~6)
and $g_\varphi$ commute with each other. 
And there exists a dense subset ${\cal F}_{g\sigma\tau}^a$,
where, for any $A\in {\cal F}_{g\sigma\tau}^a$, 
$\tau_z(A)$, $\sigma^{(+)}_z(A)$ and $g_z(A)$ are analytic on 
the whole complex
plane of $z$ (cf. the arguments of
Proposition 2.5.22 of Ref.~\citen{Bratteli}). 
Then, we have

\medskip

\begin{quote}
\noindent{\bf Corollary 7}: \ Let (i) ${\hat I}\in {\cal F}_{g\sigma\tau}^a
\cap D(\hat\delta_\xi^{(N)})\cap D(\hat\delta_\xi^{(E)})$ 
be gauge-invariant:
$g_\varphi({\hat I})={\hat I}$, (ii) $W\in D(\hat\delta_\xi^{(N)})\cap 
D(\hat\delta_\xi^{(E)})$ and 
\begin{eqnarray}
&&\int_{-\infty}^{+\infty} dt |\langle 
\tau_t(\delta{\hat I})\delta{\hat I}\rangle_+|<+\infty \ , 
\label{Coro7_integrablity}\\
&&\int_{-\infty}^{+\infty} dt \Vert [\tau_t(\hat\delta_\xi^{(\lambda)}
(W),\delta\hat I]\Vert <+\infty \ \ (\lambda=N, E) ,
\label{Coro7_L1}
\end{eqnarray}
where $\delta\hat I\equiv \hat I-\langle \hat I\rangle_+$ stands for 
the fluctuation of $\hat I$ and $\hat\delta_\xi^{(\lambda)}$ 
($\lambda=N,E$) are the infinitesimal generators of $\xi_s^{(\lambda)}$
($\lambda=N,E$) formally expressed as 
$\xi_s^{(N)}(A)=e^{i(N_L-N_R)s}Ae^{-i(N_L-N_R)s}$ and
$\xi_s^{(E)}(A)=e^{i(H_L-H_R)s}Ae^{-i(H_L-H_R)s}$ (i.e.,
$\hat\delta_\xi^{(\lambda)}(A)={d\over ds}\xi_s^{(\lambda)}(A)|_{s=0}$).
Consider the Fourier transform $S_I(\omega)$ of 
the symmetrized correlation function
\begin{equation}
S_I(\omega)\equiv {1\over 2}\int_{-\infty}^{+\infty} dt e^{i\omega t}
\langle\{ \tau_t(\delta{\hat I})
\ \delta{\hat I}
+\delta{\hat I} \ \tau_t(\delta{\hat I})
\}\rangle_+ \ ,
\end{equation}
then
\begin{eqnarray}
&&(e^{\bar\beta\hbar\omega}-1)S_I(\omega)-(e^{\bar\beta\hbar\omega}+1)
{\rm Re}(\hbar\omega G(\omega))\cr
&&=
{\Delta\beta\over 2}
\int_{-\infty}^{+\infty} dt e^{i\omega t}
\int_0^1 dv \Big\langle \tau_t(\delta\hat I)
\xi_{iv}^{(+)}
\Bigl(
i\hat\delta_\xi^{(E)}(\delta\hat I)-\int_{-\infty}^0{dt'\over\hbar}
[\tau_{t'}(\hat\delta_\xi^{(E)}(W)),\delta\hat I]
\Bigr)
\Big\rangle_+
\cr
&&-
{\Delta\aleph\over 2}
\int_{-\infty}^{+\infty} dt e^{i\omega t}
\int_0^1 dv \Big\langle \tau_t(\delta\hat I)
\xi_{iv}^{(+)}
\Bigl(i
\hat\delta_\xi^{(N)}(\delta\hat I)-\int_{-\infty}^0{dt'\over\hbar}
[\tau_{t'}(\hat\delta_\xi^{(N)}(W)),\delta\hat I]
\Bigr)
\Big\rangle_+ \ , \cr
&& \label{NonEqFDT}
\end{eqnarray}
where $\xi_s^{(+)}$ is the map 
$\xi_s^{(+)}=\sigma_s^{(+)}\tau_{-\bar\beta\hbar s}g_{\bar\aleph s}$,
$\bar\beta=(\beta_L+\beta_R)/2$ the average inverse temperature,
$\Delta\beta=\beta_L-\beta_R$ the difference of
inverse temperatures, 
$\Delta\aleph=\beta_L\mu_L-\beta_R\mu_R$
the affinity difference
and 
$\bar\aleph=(\beta_L\mu_L+\beta_R\mu_R)/2$
the average affinity.
Note that the infinitesimal generator
$\hat\delta_\xi^+$ defined by $\hat\delta_\xi^+(A)\equiv{d\over ds}
\xi_{s}^{(+)}(A)|_{s=0}$ ($A\in D(\hat\delta_\xi^+)$)
is of order of $\Delta\beta$ and $\Delta\aleph$.
\end{quote}

\medskip

\noindent
From the arguments of the previous section, we have 
$\hat\delta_\xi^{(N)}(W)=\hbar(J_R-J_L)$ and
$\hat\delta_\xi^{(E)}(W)=\hbar(J_R^E-J_L^E)$,
and, thus, the terms involving $t'$-integrals are 
correlation functions among three current operators.
On the contrary, the terms 
$i\hat\delta_\xi^{(\lambda)}(\delta\hat I)$
\break
($\lambda=N,E$) depend on the interaction $W$
and, in general, do not have simple physical meaning.
For the model of an AB ring with a dot,
when $w=0$,
$\beta_L=\beta_R=\beta$ and the left-hand side of 
(\ref{NonEqFDT}) is absolutely integrable,
the response and correlation functions with respect to
the average current:
$\hat I=-e(J_R-J_L)/2$ satisfy

\begin{eqnarray}
&&\int_{-\infty}^{+\infty}{d\omega\over 2\pi}
e^{-i\omega t}
\left\{(e^{\beta\hbar\omega}-1)S_I(\omega)-(e^{\beta\hbar\omega}+1)
{\rm Re}(\hbar\omega G(\omega))\right\}
\cr
&&=
{\beta\Delta\mu\over 2i}
\int_0^1 dv \Big\langle \tau_t(\delta\hat I)
\xi_{iv}^{(+)}
\Bigl({e\over 2\hbar}W
+{2 \over ie}\int_{-\infty}^0dt'
[\tau_{t'}(\delta\hat I),\delta\hat I]
\Bigr)
\Big\rangle_+ \cr
&&=
{\beta\Delta\mu\over 2i}\Big\{
{e\over 2\hbar}\langle \tau_t(\delta\hat I)W\rangle_+
+{2 \over ie}\int_{-\infty}^0dt'
\Bigl\langle \tau_t(\delta\hat I)
[\tau_{t'}(\delta\hat I),\delta\hat I]
\Bigr\rangle_+\Big\}+{\rm O}(\Delta\mu^2) \nonumber 
\end{eqnarray}
where $\Delta\mu=\mu_L-\mu_R$ is the chemical potential difference.
Thus, the imperfection of the fluctuation dissipation relation
is equal to a sum of two correlation functions, the one between 
the current and the interaction $W$ and the other
among three currents.
We believe that this relation is a first step
towards a quantum analog to the equality between the violation
of fluctuation-dissipation relation and energy dissipation
obtained for certain classical systems 
by Harada-Sasa\cite{sasa1} and Teramoto-Sasa\cite{sasa2}
or to a nonequlibrium extension of fluctuation-dissipation
relation derived for classical Langevin systems
by Speck and Seifert\cite{Seifert}.

\medskip

\noindent{\it Proof of Corollary 7}: Integrability 
(\ref{Coro7_integrablity}) guarantees 
the existence of $S_I(\omega)$ and $\hbar\omega G(\omega)$.
By integrating an analytic function
$e^{i\omega t}\langle \delta\hat I\tau_t(\delta\hat I)\rangle_+$ on
a rectangle\break
$[-T_1, T_2]\cup [T_2,T_2+i\hbar \bar\beta]
\cup [T_2+i\hbar \bar\beta,-T_1+i\hbar \bar\beta]
\cup [-T_1+i\hbar \bar\beta,-T_1]$ in the complex $t$-plane
with $[z_1,z_2]$ a segment starting from $z_1$ and terminating at $z_2$, 
and by taking the limit of $T_1, T_2\to +\infty$, 
we obtain
\begin{equation}
\int_{-\infty}^{+\infty
} dt e^{i\omega t}
\langle \delta\hat I \tau_t(\delta\hat I)\rangle_+
=
e^{-\hbar\bar\beta \omega}\int_{-\infty}^{+\infty
} dt e^{i\omega t}
\langle \tau_{-i\hbar\bar\beta}(\delta\hat I)
\tau_t(\delta\hat I)\rangle_+ \ .\label{CorP1}
\end{equation}
Note that the intergals on the segments $[T_2,T_2+i\hbar \bar\beta]$
and $[-T_1+i\hbar \bar\beta,-T_1]$ vanish for $T_1,T_2\to +\infty$
because of the mixing property given by Proposition~2.
On the other hand, Proposition~6 and the gauge invariance of $\hat I$:
$g_{i\bar\aleph}(\hat I)=\hat I$ imply
\begin{eqnarray}
&&\langle \tau_{-i\hbar\bar\beta}(\delta\hat I)\tau_t(\delta\hat I)\rangle_+ =
\langle \tau_{-i\hbar\bar\beta}(g_{i\bar\aleph}(\delta\hat I))
\tau_t(\delta\hat I)\rangle_+ =
\langle \tau_t(\delta\hat I)
\sigma_i^{(+)}(
\tau_{-i\hbar\bar\beta}(g_{i\bar\aleph}(\delta\hat I)))\rangle_+ \cr
&&=
\langle \tau_t(\delta\hat I)
\xi_i^{(+)}(\delta\hat I)\rangle_+ 
=\langle \tau_t(\delta\hat I)
\delta\hat I\rangle_+ 
+i\int_0^1 dv \langle \tau_t(\delta\hat I)
\xi_{iv}^{(+)}(\hat\delta_\xi^+(
\delta\hat I))\rangle_+ \ , \label{CorP2}
\end{eqnarray}
where $\delta\hat I\in D(\hat\delta_\xi^+)$ is
shown later. 
It is easy to see $\xi_s^{(+)}=\gamma_+^{-1}\xi_s^{(0)}
\gamma_+$ where\break
$\xi_s^{(0)}=\sigma_s \tau^{(0)}_{-\hbar\beta s}g_{\bar\aleph s}$,
and 
$\xi_s^{(0)}(A)=\xi_{\Delta \bar\beta s/2}^{(E)}
\bigl(\xi_{-\Delta \bar\aleph s/2}^{(N)}(e^{-i(\bar\beta\epsilon_0-
\bar\aleph)c^\dag c s}Ae^{i(\bar\beta\epsilon_0-
\bar\aleph)c^\dag c s})\bigr)$.
Thus,
$\xi_s^{(+)}(\delta\hat I)=\gamma_+^{-1}\Bigl(
\xi_{\Delta\beta s/2}^{(E)}
\bigl(\xi_{-\Delta\aleph s/2}^{(N)}(
\gamma_+(\delta\hat I) )\bigr)\Bigr)$.
Since $\xi_s^{(+)}(\delta\hat I)$ has the same structure as 
$\sigma_s^{(+)}(A)$, one can show $\delta\hat I\in D(\hat\delta_\xi^+)$ and
\begin{eqnarray}
\hat\delta_\xi^+(\delta\hat I)&=&{\Delta\beta\over 2}
\bigl\{
\hat\delta_\xi^{(E)}(\delta\hat I)+{i\over \hbar}\int_{-\infty}^0dt
[\tau_t(\hat\delta_\xi^{(E)}(W)),\delta\hat I]
\bigr\}\cr
&&-
{\Delta\aleph\over 2}\bigl\{
\hat\delta_\xi^{(N)}(\delta\hat I)+{i\over\hbar}\int_{-\infty}^0dt
[\tau_t(\hat\delta_\xi^{(N)}(W)),\delta\hat I]
\bigr\} \ , \label{CorP3}
\end{eqnarray}
from the conditions (i)-(ii) and (\ref{Coro7_L1}) as in
the proof of Proposition~6. 
Combining (\ref{CorP1}), (\ref{CorP2}), (\ref{CorP3}),
and using 
$S_I(\omega)-{\rm Re}(\hbar \omega G(\omega))=\int_{
{\bf R}} dt e^{i\omega t}
\langle \delta\hat I \tau_t(\delta\hat I)\rangle_+$
and 
$S_I(\omega)+{\rm Re}(\hbar \omega G(\omega))=\int_{
{\bf R}} dt e^{i\omega t}
\langle\tau_t(\delta\hat I)\delta\hat I\rangle_+$,
we obtain the desired result. \ ({\it Q.E.D})

\section{Conclusions}
As pointed out by Ruelle\cite{Ruelle1} and clearly seen from Proposition~1, nonequlibrium 
steady states investigated 
so far are constructed through the scattering approach. In this sense, 
the present approach can
be regarded as an extension of Landauer-B\"uttiker's approach\cite{LandauerButtiker} to 
electronic transports 
in mesoscopic systems.  And there exist a class of mesoscopic systems 
to which the present formalism
is applicable, such as the Ahoronov-Bohm 
ring with a quantum dot. 
Since we have a full characterization of NESS for
non-interacting systems, one may develop approximations 
such as the mean-field approximation\cite{TakahashiTasaki2}.
These aspects will be discussed elsewhere.

Before closing this article, we look through a relation between the 
dynamical reversibility and
irreversible evolution towards a steady state in the sense of
Proposition~1~\cite{STMatsui}. 
We assume that the dynamics is symmetric with respect to
a time reversal operation, namely, there exists an antilinear 
operation $A\to \iota(A)$ such that 
$\iota(AB)=\iota(A)\iota(B), \ \iota(\alpha A+B)=\alpha^*\iota(A)+\iota(B), \  \iota^2(A)=A$ 
and $\iota(\tau_t(\iota (A)))=\tau_{-t}(A)$ ($A,B\in {\cal F}, 
\ \alpha\in {\bf C}$).
The time reversal operation on a state $\langle \cdots \rangle$ 
is, then, defined by 
$\langle A \rangle^{\rm TR}\equiv \langle \iota(A^\dag) \rangle$. 
For the present model, one can choose $\iota(a_r(f))=\int dk f(-k)a_{kr}$, 
$\iota(c)=c$. Then,
when $\iota(W)=W$ (when $\varphi=0$ for the model of an Ahoronov-Bohm ring with 
a dot),
the system has a time-reversal symmetry. 
Under the
assumption that the initial state is time-reversal symmetric 
$ \langle A \rangle_{loc}^{\rm TR}= 
\langle A\rangle_{loc}$, 
let us carry out the following thought experiment: 
(i) \ Let the system autonomously evolve up to $t=t_0$. 
(ii) \ The time reversal operation
is performed at $t=t_0$. And (iii) \ let the system autonomously 
evolve once again.
Just before the time reversal operation, the system is in the state $\langle A\rangle_{t_0-}=
\langle \tau_{t_0}(A)\rangle_{loc}$ and, just after the time reversal operation, the state
becomes
$$
\langle A \rangle_{t_0+}=
\langle A \rangle_{t_0-}^{\rm TR} = 
\langle \tau_{t_0}(\iota(A^\dag)) \rangle_{loc} =
\langle \iota(\tau_{-t_0}(A)^\dag) \rangle_{loc}=
\langle \tau_{-t_0}(A) \rangle_{loc}
\ ,
$$
which evolves further as 
$\langle A \rangle_{t}=\langle \tau_{t-t_0}(A) \rangle_{loc}$.
Then, the system comes back at $t=2t_0$ to the state just before 
the time reversal
operation, as expected. Note that, if $t_0>0$ is large enough, the state $\langle A \rangle_{t_0-}$
just before the time reversal operation is close to the steady state $\langle A \rangle_+$,
while the state $\langle A \rangle_{t_0+}$ just after the time reversal operation
is close to another steady state $\langle A\rangle_-$. In other words, 
the time reversal operation
discontinuously derives the system from a state close to 
$\langle\cdots\rangle_+$ to the one close to
$\langle\cdots\rangle_-$. On the other hand, by the `natural' evolution $\tau_t$, the system 
changes towards the steady state $\langle\cdots\rangle_+$. Hence, the dynamical 
reversibility is fully consistent with irreversible state evolution 
in the sense of
Proposition~1\cite{STMatsui}.

It is interesting to revisit Loschmidt's criticism to the work of 
Boltzmann\cite{Ehrenfest}.
Although the dynamical reversibility is consistent with the irreversible 
state evolution, 
Loschmidt's criticism can be applied to the relative entropy 
$S(\rho_{loc}|\rho_t)$.
Indeed, in the above thought experiment, let $\langle J_S\rangle_t$ be 
the entropy production
at time $t$, then, when $t$ is slightly larger than $t_0$, 
$\langle J_S\rangle_t$ is close to
$\langle J_S\rangle_-(<0$\footnote{This 
follows from (\ref{EnPro}) and l'Hospital's
rule as the proof of $\langle J_S\rangle_+\ge 0$.})
and should itself be negative.
Thus, because of (\ref{EnPro}),
there is a period when the relative entropy $S(\rho_{loc}|\rho_t)$
decreases. Contrary to Boltzmann's reply\cite{Ehrenfest}, 
these states are typical in the sense that
they evolve towards the steady state $\langle\cdots\rangle_+$ for 
$t\to\infty$.
In other words, Loschmidt's criticism does not deny the consistency of irreversible 
phenomena with dynamical reversibility 
and it just 
shows that the relative entropy is not an appropriate 
thermodynamic entropy 
for general cases. 
Another criticism to Boltzmann's work by Zermelo\cite{Ehrenfest} is 
not applicable to 
the 
present case since
the recurrence time is infinitely long as a result of the 
infinite extension of the system.

From the point of view of the second law of thermodynamics,
one may be satisfied with all these features, particularly the properties of the entropy 
production discussed in Sec.~3. However, one should remind that the `correct' form
of the entropy production is obtained {\it because} we start from a local 
equilibrium
state where each subsystem is in a canonical state.
As the canonical states are very outcome of the second law, the present results do
not give a dynamical proof of the second law, but show that, once one starts 
from canonical ensembles or their combinations, the dynamics derives the system consistently with  
the second law of thermodynamics. 

\section*{Acknowledgments}
The authors thank Professors L. Accardi, P. Facchi, P. Gaspard, 
H. Hayakawa, C. Jarzynski, 
T. Matsui, H. Nakazato, I. Ohba, M. Ohya, I.~Ojima, S. Pascazio,
S. Sasa, K.~Sekimoto A.~Shimizu, S. Takesue, H. Tasaki, K. Yuasa 
for discussions
and comments.
One of them (ST) is grateful to Professors T. Ohta and M. Murase
for their hospitality and support at ``International Symposium on Physics of 
Non-Equilibrium Systems'', (3-7 Oct. 2005, Nishinomiya \& Kyoto).
This work is partially supported by Grant-in-Aid for Scientific 
Research (B) (No.17340114) and (C) (No.17540365) 
from the Japan Society of the Promotion of Science, 
as well as
by a Grant-in-Aid for Scientific Research of Priority Areas 
``Control of Molecules in Intense Laser Fields'' (No.14077219), 
the 21st Century COE Program at Waseda University ``Holistic Research 
and Education Center for Physics of Self-organization Systems''
and ``Academic Frontier'' Project 
from the Ministry of Education, Culture, Sports, Science and 
Technology of Japan.

\end{document}